\documentclass[a4paper, 10 pt, conference]{IEEEtran}  
\usepackage{enumitem}
\usepackage{xcolor}
\usepackage{url}
\usepackage{amsmath}
\usepackage{balance}
\IEEEoverridecommandlockouts                              

\usepackage[doi=false,isbn=false,url=false,eprint=false]{biblatex}
\DeclareFieldFormat{urldate}{}

\title{\LARGE
Homophily in Complex Networks: Measures, Models, and Applications
}

\author{
    \IEEEauthorblockN{Akrati Saxena\IEEEauthorrefmark{1}, Gaurav Kumar\IEEEauthorrefmark{2} and Chandrakala Meena\IEEEauthorrefmark{2}}
    \IEEEauthorblockA{\IEEEauthorrefmark{1}Leiden Institute of Advanced Computer Science, Leiden University, The Netherlands\\ a.saxena@liacs.leidenuniv.nl}
        \IEEEauthorblockA{\IEEEauthorrefmark{2}Indian Institute of Science Education and Research (IISER) Pune, India }
}

\begin{document}

\maketitle

\begin{abstract}

Homophily, the tendency of individuals to connect with others who share similar attributes, is a defining feature of social networks. Understanding how groups interact, both within and across, is crucial for uncovering the dynamics of network evolution and the emergence of structural inequalities in these network. This tutorial offers a comprehensive overview of homophily, covering its various definitions, key properties, and the limitations of widely used metrics. Extending beyond traditional pairwise interactions, we will discuss homophily in higher-order network structures such as hypergraphs and simplicial complexes. We will further discuss network generating models capable of producing different types of homophilic networks with tunable levels of homophily and highlight their relevance in real-world contexts. The tutorial concludes with a discussion of open challenges, emerging directions, and opportunities for further research in this area.

\end{abstract}

\section{Target Audience and Prerequisites} 

This tutorial is intended for researchers in the fields of social and complex networks who are interested in understanding network structure, advancing homophily measures, and exploring generative models. It will be particularly relevant to those studying systems where group interactions play a central role, such as social, organizational, and communication networks. Participants should have prior exposure to basic concepts in graph theory or network science. The tutorial aims to serve as a valuable resource for researchers investigating network evolution, structural inequalities, group formation, and the development of network-based models and metrics.

\section{Related Past Tutorials} 

Previous tutorials, such as those by Venkatasubramanian et al. \cite{venkatasubramanian2021fairness, venkatasubramanianFairnessNetworksSocial2021} on fairness in networks had a brief discussion of homophily, as their primary focus was on algorithmic fairness. To the best of our knowledge, no dedicated tutorials focusing on homophily, especially its definitions, measurement, and modeling in both pairwise and higher-order network structures, have been presented at ASONAM or other major conferences.

\section{Other Tutorials}

Dr. Akrati Saxena has organized and presented following tutorials in the past:

\begin{itemize}
    \item Roles Analytics in Networks - Foundations, Methods and Applications tutorial at ICDM 2021 conference (More details: \url{https://cswzhang.github.io/icdm-tutorial-2021/})
    \item Network Science Applications to Education in the 21st Century tutorial at ASONAM 2021 conference \cite{geranetwork} (More details: \url{https://sites.google.com/view/asonam-tutorial-2021})
\end{itemize}

\section{Tutors}

\textbf{Dr. Akrati Saxena} is an Assistant Professor at the Leiden Institute of Advanced Computer Science (LIACS), Leiden University, Netherlands, and an Adjunct Professor at the University of Victoria, Canada. Previously, she worked as a Research Fellow at Eindhoven University of Technology, Netherlands, and the National University of Singapore. Dr. Saxena serves as an Associate Editor for Social Network Analysis and Mining and PLOS Complex Systems journals. Her research interests span social network analysis, complex networks, computational social science, social media, and fairness. Her current research focuses on developing fairness measures and designing fair algorithms for network analysis.

\textbf{Mr. Gaurav Kumar} is a Junior Research Fellow at Department of Physics, Indian Institute of Science Education and Research (IISER) Pune, India. He received his Master's degree in Physics from Indian Institute of Technology (IIT) Gandhinagar, India. His research interests include social network analysis, homophily and contagion models on networks.

\textbf{Dr. Chandrakala Meena} is an assistant professor at Indian Institute of Science Education and Research (IISER) Pune, India. She received her Ph.D. in Physics from IISER Mohali, India under supervision of Prof. Sudeshna Sinha. Her research broadly focuses on the dynamical behaviour and pattern formation in nonlinear systems and complex networks. 

\section{Tutorial Outline}
The total presentation time is 2.5-3 hours. Outline is as follows.
\begin{enumerate}[label=\arabic*.]
    \item Introduction and Motivation [20 min]
    \begin{itemize}[label=-]
        \item Importance of understanding group interactions in networks
        \item Homophily as a central concept in social network analysis
        \item Structural inequalities in networks \cite{saxenaFairSNAAlgorithmicFairness2024}
        \item Goals of the tutorial
    \end{itemize}
    
    \item Homophily in Networks [30 min]
    \begin{itemize}[label=-]
        \item Classical definitions: assortativity, Coleman index, edge-based measures
        \item Properties of a Robust Homophily Measure: Baselines, Interpretability, and Sensitivity 
    \end{itemize}
    
    \item Homophily Beyond Pairwise Interactions [30 min]
    \begin{itemize}[label=-]
        \item Hypergraphs and Simplicial Complexes
        \item How higher-order structures capture group interactions
        \item Extending homophily to higher-order interactions
    \end{itemize}
    
    \item Generating Networks with Homophily [25 min]
    \begin{itemize}[label=-]
        \item Stochastic block models and variants
        \item Generating graphs and hypergraphs with tunable homophily
        \item Simulation examples and parameter tuning
    \end{itemize}
    
    \item Applications and Empirical Studies [25 min]
    \begin{itemize}[label=-]
        \item Homophily in political networks, co-authorship, and online platforms
        \item Homophily and inequality, polarization, echo chambers
        \item Impact of homophily on Algorithmic Fairness 
    \end{itemize}
    
    \item Conclusion [20 min]
    \begin{itemize}[label=-]
        \item Conclusion
        \item Open discussion about future directions
    \end{itemize}
\end{enumerate}
\subsection*{Any specific audio/video/computer requirements}
We do not have any specific requirement.

\subsection{Introduction and Motivation}

We will introduce homophily as a core principle in social and complex networks, emphasizing its role in shaping group formation, inequality, and access to social capital. Starting from classical definitions such as Coleman’s index and assortativity, we will highlight recent advances that address their limitations and extend homophily to higher-order structures. 
By setting this foundation, the introduction will prepare participants to appreciate the methodological, theoretical, and application-oriented aspects of homophily that will be explored throughout the tutorial.

\subsection{Homophily in Networks}

In this part, we will present key metrics used to quantify homophily in complex networks, including the measures mentioned below.
\subsubsection{Edge Homophily}
The most straightforward measure, it is simply the fraction of edges in the network that connect nodes with the same community label \cite{mironovRevisitingGraphHomophily2024}. Its expected value in a random graph is highly dependent on the relative sizes of the communities, making it unreliable for comparing networks with different compositions.

\subsubsection{Node Homophily}
This measure calculates the average fraction of a node's neighbors that belong to the same community \cite{mironovRevisitingGraphHomophily2024}.

\subsubsection{Class Homophily}
This measure aggregates homophily at the class level by comparing the fraction of a group's total edge connections that are internal to what would be expected by random chance \cite{mironovRevisitingGraphHomophily2024}. 

\subsubsection{E-I Index}
A simple count of within-group (Internal) versus between-group (External) links, it ranges from -1 (perfect homophily) to +1 (perfect heterophily) \cite{krackhardtInformalNetworksOrganizational1988}. Its main drawback is that it ignores the opportunity structure created by group sizes; a score of 0 does not mean random mixing.

\subsubsection{Coleman's Homophily Index}
One of the earliest measures to formally account for the opportunity structure, it compares the observed number of within-group choices to the number expected by chance \cite{colemanRelationalAnalysisStudy1958}. It is calculated for each group separately, allowing for asymmetric results.

\subsubsection{Freeman's Segregation Index}
This measure conceptualizes segregation as the restriction of ties between groups, comparing the observed proportion of between-group ties to a random baseline \cite{freemanSegregationSocialNetworks1978,bojanowskiMeasuringSegregationSocial2014}. 

\subsubsection{Odds-Ratio for Within-Group Ties (ORWG)}
The ORWG quantifies homophily by comparing the likelihood of a tie between two nodes belonging to the same group with the likelihood of a tie between nodes from different groups \cite{moodyRaceSchoolIntegration2001, bojanowskiMeasuringSegregationSocial2014}. A notable advantage of this measure is its independence from the marginal distribution of group sizes, making it suitable for networks with imbalanced group structures. However, a key limitation is its high sensitivity to isolates, which can distort the assessment of homophily in sparse or fragmented networks.

\subsubsection{Nominal Assortativity Coefficient}
A widely used measure based on the network's mixing matrix \cite{newmanMixingPatternsNetworks2003}. It is severely biased by group size imbalance and cannot capture asymmetric mixing patterns \cite{karimiInadequacyNominalAssortativity2023}.

\subsubsection{Gupta, Anderson, and May's Q}
An early measure of ``within-group mixing'' \cite{guptaNetworksSexualContacts1989}. Its main drawback is that it gives misleading results with unequal group sizes because it weights each type of vertex equally, giving disproportionate weight to small groups \cite{newmanMixingPatternsNetworks2003}.

\subsubsection{Adjusted Nominal Assortativity}
The adjusted nominal assortativity addresses the group-size bias of Newman’s $r$ by normalizing mixing matrix elements with respect to group fractions \cite{karimiInadequacyNominalAssortativity2023}. This adjustment enables accurate estimation of assortativity, independent of group size imbalance.

\subsubsection{Unbiased Homophily}
Unbiased homophily is designed to satisfy key theoretical properties such as monotonicity and constant baseline \cite{mironovRevisitingGraphHomophily2024}, ensuring reliability for cross-dataset comparisons.  

\subsubsection{Segregation Matrix Index (SMI)}
The SMI evaluates group cohesiveness by comparing the density of internal and external ties \cite{fershtmanCohesiveGroupDetection1997}. It is computed separately for each group, enabling asymmetric assessments.

\subsubsection{Spectral Segregation Index (SSI)}
The SSI quantifies segregation under the principle that an individual is more segregated if their neighbors are also segregated \cite{echeniqueMeasureSegregationBased2007}. For each group, SSI is defined as the largest eigenvalue of its row-normalized within-group interaction matrix, and group scores are aggregated as a size-weighted average of these eigenvalues.

\subsubsection{Degree-Weighted Homophily (DWH)}
Degree-Weighted Homophily \cite{golubHowHomophilyAffects2012} measures homophily by testing all possible ways to split the communities into two super-groups and finding the split where within-group connections are most concentrated compared to between-group connections. That is, DWH identifies the most segregated division in the network.

\subsubsection{Popularity-Homophily Index}
The Popularity-Homophily Index extends homophily measurement to directed graphs by weighting ties according to the popularity (e.g., PageRank) of the recipient node \cite{oswalPopularityHomophilyIndexNew2021}. Since its absolute magnitude can be misleading, it is most suitable for relative comparisons.

\subsubsection{The Random Coloring Model (Z-Score)}
It provides a statistical test for the significance of observed homophily by comparing against a null model in which node labels are randomly permuted on the fixed graph structure \cite{apollonioNovelMethodAssessing2022}.

\subsection{Homophily Beyond Pairwise Interactions}
\subsubsection{Affinity Score}  
In a $k$-uniform hypergraph, the affinity score for a given class measures the tendency of its members to form groups containing exactly $t$ nodes from the same class. This observed tendency (the ``type-$t$ affinity score'') is evaluated against a random baseline, either as a ratio or as a normalized difference, to determine whether it is higher or lower than expected by chance. This enable the characterization of patterns including simple homophily (preference for fully uniform groups), majority homophily (preference for groups where a class is dominant), and monotonic homophily (increasing preference as class representation grows) \cite{veldtCombinatorialCharacterizationsImpossibilities2023}.  

\subsubsection{Simplicial Homophily Score}
Simplicial homophily measures whether groups of size $k$ in a simplicial complex are more likely to consist of nodes of the same type than expected from the lower-order structure. The affinity score is the fraction of observed $k$-simplices whose nodes all share the same feature, and the baseline is the same fraction computed over all possible $k$-simplices that could exist given the $(k-2)$-skeleton. The ratio of these two values defines the simplicial homophily score \cite{sarkerHigherorderHomophilySimplicial2024}.

\subsubsection{Hyperedge Homophily}  
It first computes the fraction of same-labeled node pairs among all possible pairs for each hyperedge, and then averages this value across all hyperedges to obtain a network-level score \cite{liWhenHypergraphMeets2025}.

\subsubsection{Node Homophily in Hypergraphs}
This measure focuses on the individual's perspective. For each node, it first computes the average proportion of same-labeled members across all groups it belongs to, and then averages this value over all nodes in the network \cite{liWhenHypergraphMeets2025}.

\subsubsection{Message Passing and $\Delta$-Homophily}  
Message Passing Homophily \cite{telyatnikovHypergraphNeuralNetworks2023} is computed through iterative message passing, where hyperedge scores are obtained by aggregating node labels and node scores by aggregating the scores of their incident hyperedges. This process computes homophily values at multiple neighborhood resolutions. The change in a node’s homophily between consecutive steps, termed $\Delta$-Homophily, captures the stability of its structural environment, with smaller changes indicating stronger homophily.  

\subsubsection{Clique-Expanded Homophily}  
We will cover Clique-expanded homophily measures \cite{telyatnikovHypergraphNeuralNetworks2023, liWhenHypergraphMeets2025}, which compute homophily by converting each hyperedge into a clique and applying standard graph-based metrics, and their drawbacks. 

\subsection{Generating Networks with Homophily}
In this part, we cover network generating models that can generate synthetic networks with homophily where homophily can be controlled using the model hyperparameter. We will cover models, including Homophily BA Model \cite{karimi2018homophily, lee2019homophily}, Diversified Homophily BA model \cite{wang2021information}, Directed Homophily Network \cite{anwar2021balanced}, Organic Growth Model \cite{stoica2018algorithmic}, HICH-BA Model \cite{saxena2023fairness}, and Homophilic Clique network model \cite{rizi2024homophily}.

\subsection{Applications and Empirical Studies}

Here, we will cover empirical studies that highlight the role of homophily in shaping diverse real-world networks, such as political, social media, and collaboration networks \cite{esteve2022homophily, macedo2024gender, zhang2018understanding}, while also revealing its contribution to structural inequalities. We will then discuss how homophily influences the algorithmic fairness of network analysis tasks, including link prediction \cite{saxena2022hm, khajehnejad2021crosswalk, saxena2022nodesim}, community detection \cite{panayiotou2024fair, de2025quantifying, de2024group}, influence maximization \cite{stoica2019fairness, stoica2020seeding}, and influence blocking \cite{saxena2023fairness}. For instance, strong within-group connectivity can lead algorithms to reinforce existing segregation, restrict information access, or disproportionately favor majority groups, thereby amplifying inequalities. Through case studies and applications, we will demonstrate that systematically measuring and accounting for homophily is crucial for developing fair and reliable algorithms across networks with varying structural and demographic characteristics.

\subsection{Conclusion and Future Directions}
We will conclude by synthesizing the key insights and outlining prospective research directions, with particular emphasis on identifying critical gaps and strategies to address them.

\printbibliography
\end{document}